\documentclass[aps,prc,showpacs,showkeys,twocolumn,superscriptaddress,groupedaddress]{revtex4-1}

\usepackage[english]{babel}
\usepackage{graphics}
\usepackage{graphicx}
\usepackage{bm}
\usepackage{verbatim}
\usepackage{amssymb}
\usepackage{braket}
\usepackage{color}
\usepackage{wrapfig}
\usepackage{pifont}
\usepackage{amsfonts}
\usepackage{mathrsfs}
\usepackage{slashed}
\usepackage{amsmath}
\usepackage{epsfig}
\usepackage{latexsym}
\usepackage{psfrag}
\usepackage{nicefrac}
\usepackage{ragged2e}
\usepackage{etoolbox}
\usepackage{tikz}
\usepackage{pgfpages}
\usepackage{dsfont}
\usepackage{mathtools}
\usepackage{bbm}
\def\n4lo{$\mathrm{N}^4\mathrm{LO}$}
\def\n3lo{$\mathrm{N}^3\mathrm{LO}$}

     \def\chiral4lo{$\mathrm{N}^4\mathrm{LO}$}

\begin{document}

\noindent
\title{Elastic proton scattering off non-zero spin nuclei}

\author{Matteo Vorabbi$^{1}$}
\author{Michael Gennari$^{2,3}$}
\author{Paolo Finelli$^{4}$}
\author{Carlotta Giusti$^{5}$}
\author{Petr Navr\'{a}til$^{2,3}$}
\author{Ruprecht Machleidt$^{6}$}

\affiliation{$~^{1}$National Nuclear Data Center, Bldg. 817, Brookhaven National Laboratory, Upton, NY 11973-5000, USA
}

\affiliation{$~^{2}$University of Victoria, 3800 Finnerty Road, Victoria, British Columbia V8P 5C2, Canada
}

\affiliation{$~^{3}$TRIUMF, 4004 Wesbrook Mall, Vancouver, British Columbia, V6T 2A3, Canada
}

\affiliation{$~^{4}$Dipartimento di Fisica e Astronomia, 
Universit\`{a} degli Studi di Bologna and \\
INFN, Sezione di Bologna, Via Irnerio 46, I-40126 Bologna, Italy
}

\affiliation{$~^{5}$Dipartimento di Fisica,  
Universit\`a degli Studi di Pavia and \\
INFN, Sezione di Pavia,  Via A. Bassi 6, I-27100 Pavia, Italy
}

\affiliation{$~^{6}$Department of Physics, University of Idaho, Moscow, Idaho 83844, USA
}

\date{\today}

%%%%%%%%%%%%%%%%%%%%%%%%%%%%%%%%%%%%%%%%%%%%%%%%%%%%%%%

\begin{abstract} 

\noindent
{\bf Background: } In recent years, we constructed a microscopic optical potential (OP) for elastic nucleon-nucleus ({\it NA}) scattering using modern approaches based on
chiral theories for the nucleon-nucleon ({\it NN}) interaction.
The OP was derived at first order of the spectator expansion in Watson multiple scattering theory and its final expression was a folding
integral between the {\it NN} $t$ matrix and the nuclear density of the target. Two- and three-body forces are consistently included both in the target and in the projectile description.

\noindent
{\bf Purpose:} The purpose of this work is to apply our microscopic OP to nuclei characterized by a ground state of spin-parity quantum numbers $J^\pi \ne 0^+$.

\noindent
{\bf Methods:} We extended our formalism to include the spin of the target nucleus. The full amplitudes of the {\it NN} reaction matrix are retained in the
calculations starting from two- and three-body chiral forces.

\noindent
{\bf Results:} The microscopic OP can be applied in the energy range $100 \le E \le 350$ MeV.
We show a remarkable agreement with experimental data for the available observables and, simultaneously, provide reliable estimates for the theoretical uncertainties.

\noindent
{\bf Conclusions:} This work paves the way toward a full microscopic approach to inelastic {\it NA} scattering, showing that the derivation of optical potentials
between states with $J^\pi \ne 0^+$ is completely under control.

\end{abstract}

\pacs{24.10.Ht;24.70.+s;25.40.Cm;11.10.Ef}

\maketitle

%%%%%%%%%%%%%%%%%%%%%%%%% Introduction %%%%%%%%%%%%%%%%%%%%%%%%%

\section{Introduction}
The optical potential (OP) is a fundamental ingredient not only in the description of elastic diffusion but also in the analysis of more complicated reactions, where it acts as input for theoretical calculations based on the distorted wave Born approximation and coupled channel methods \cite{FESHBACH1958357, Glendenning, PaetzSchieck}. 

The study of the OP within the framework of microscopic approaches \cite{DICKHOFF2019252, Amos_review} provides, in our opinion, multiple sources of scientific interests. Even if it is true that a phenomenological approach is generally preferred to achieve a more accurate description of the available experimental data, nowadays, with the upcoming facilities for exotic nuclei (FAIR at GSI \cite{doi:10.1080/10619127.2018.1495476},  SPIRAL2 at GANIL \cite{doi:10.1142/9789814632041_0003} or SPES at LNL \cite{doi:10.1142/9789814632041_0009}, just to mention some of the most important projects), 
we strongly believe that a microscopic approach to the OP would be the preferred way to make reliable predictions and to assess the impact of unavoidable approximations.

Interest in OPs has been renewed in the scientific community over the last few years and several works have been devoted to this topic. We mention the most relevant developments in our opinion: microscopic dispersive OPs \cite{dickhoff_2017, PhysRevC.101.044303}, OPs within a self-consistent Green's function approach \cite{Idini2017, PhysRevLett.123.092501}, OPs from Coupled-Cluster calculations \cite{PhysRevC.95.024315, PhysRevC.98.044625, rotureau_fphy.2020.00285}, chiral symmetry inspired  OPs \cite{PhysRevC.100.014601, PhysRevC.101.064613, Whitehead:2020wwb, durant_2018},  nonlocal OPs \cite{Arellano_2018,arellano_2019,Arellano_2020}, $g$-matrix calculations \cite{PhysRevC.92.024618, PhysRevC.96.059905, Kohno_ptep_pty001, PhysRevC.98.054617, PhysRevC.102.024611}, global OPs \cite{PhysRevC.99.034605, PhysRevC.100.014613}, and OPs based on No-Core Shell Model (NCSM) calculations \cite{Burrows:2018ggt, Burrows:2020qvu}.

A microscopic OP was derived from chiral nuclear potentials in a series of manuscripts we produced over the last few years, 
starting from the first work \cite{Vorabbi:2015nra}, where a microscopic OP was introduced following the well known procedure of Watson \cite{Riesenfeld:1956zza}, then followed by Refs. \cite{Vorabbi:2017rvk,Vorabbi_2018}, where the agreement with experimental data and phenomenological approaches was successfully tested. As main achievements of our latest work, it is worth mentioning the inclusion of three-body ({\it 3B}) forces \cite{Vorabbi:2020cgf}, the application to translationally invariant nonlocal densities derived within the NCSM framework \cite{Gennari:2017yez}, and the extension of our OP to antiproton-nucleus elastic scattering \cite{Vorabbi_2020}. 
We note that in all these works we never investigated nuclei with spin different from zero. This is the case we want to study with this manuscript, that can be seen as a natural follow-up of our previous work. 
To our knowledge, this is the first time that a microscopic OP derived within the Watson multiple scattering theory using NCSM nonlocal densities is applied to non-zero spin nuclei.

In regard to our latest work, the present investigation is connected to the importance of {\it 3B} contributions in nuclear observables. We recently studied the effects of such contributions in nucleon-nucleus ({\it NA}) elastic scattering observables \cite{Vorabbi:2020cgf}, demonstrating that {\it 3B} effects are relevant for quantities that depend on the spin of the nucleons. In this context, it is important to verify whether these conclusions also hold for nuclei whose ground state quantum numbers are different from $0^+$.

In recent years, experimental efforts have multiplied to develop the technologies necessary to study the elastic scattering of protons (and ions) by using inverse kinematics \cite{Chung:2015sza,EXL:2016glr,Sakaguchi:2017hoe,Chebotaryov:2018ilv,Yue:2019ftu,*Yue:2019ftu_erratum,Dobrovolsky:2019wzt,Dobrovolsky:2021ggf,Egelhof:2021smc}. This configuration is necessary to study exotic nuclei that have very short average life times.
Some recent studies have tried to use such experiments with the purpose of determining the density of matter of nuclear 
systems \cite{Chung:2015sza, Sakaguchi:2017hoe, Dobrovolsky:2019wzt, Dobrovolsky:2021ggf}. However, these measurements are subject to some criticisms and are not free from sizable uncertainties. It is hence important to establish a microscopic perspective for understanding the relation of proton elastic scattering to the density of nuclear matter.
In the analysis of these experiments an essential step is the subtraction of contributions from the inelastic channel.
In this perspective, if we wish to establish a consistent microscopic approach for inelastic {\it NA} scattering, which is our long-term goal, it is mandatory to test the microscopic OP on states with spin-parity quantum numbers  J$^\pi \ne 0^+$, which is the goal of the present work.

For example, at the GSI using the IKAR chamber, the elastic diffusion of $^{6,8}$He \cite{He68} and of $^{12,14-17}$C \cite{Dobrovolsky:2021ggf} were measured on a hydrogen target at energies near 700 MeV/nucleon. Another very interesting experiment performed at the GSI was the study of the scattering of $^{56,58}$Ni nuclei on hydrogen targets at energies near 400 MeV/nucleon in inverse kinematics for the determination of the distribution of nuclear matter \cite{LIU2020135776}.
The only theoretical approach used to analyze the experimental data is the Glauber model which contains some phenomenological input, limiting its predictive power.

In the near future many interesting experiments will be carried out at the GSI which will allow for the exploration of unknown areas of the nuclide table, of great interest to the nuclear and astrophysics communities. Among the many experiments it is worth mentioning the EXL experiment, which will investigate direct reactions of light ions in inverse kinematics \cite{EXL:2016glr,Egelhof:2021smc}.

Other facilities are in operation, for example, the CSRe storage ring of HIRFL-CSR \cite{Cai_2007} and also the RI Beam Factory (RIBF)  at RIKEN \cite{RIKEN}, which has recently proposed inverse kinematics measurements at high momentum transfer \cite{Chebotaryov:2018ilv}.
This is a very interesting aspect because in general the diffusion at low momentum is sensitive to the surface density while the internal region requires high momentum transfer.
Such processes are characterized by many-body effects that make their theoretical description an extremely hard task.

In this work we extend our previous analyses of elastic proton scattering off finite nuclei, focussing our efforts towards non-zero spin targets. 
In particular, we are interested in the following set of nuclei: $^{13}$C (with quantum numbers $J^{\pi} = {1/2}^-$),
$^{6}$Li ($J^{\pi} = {1}^+$), $^{7}$Li ($J^{\pi} = {3/2}^-$), and
$^{10}$B ($J^{\pi} = {3}^+$), for which experimental data in the energy
range $100 \, \mathrm{MeV} \le \mathrm{E} \le 300 \, \mathrm{MeV}$ are available.
In addition, we also performed calculations on $^{9}$C ($J^{\pi} = {3/2}^-$) which has been measured in inverse kinematics configuration.
This set of nuclei allows us to test 
the validity of our microscopic OP \cite{Vorabbi:2015nra, Vorabbi:2017rvk, Vorabbi_2018, Vorabbi_2020} when extended to spin-unsaturated nuclei with different values of the spin.

The main difference with respect to previous calculations on spin-zero nuclei \cite{Vorabbi:2015nra, Vorabbi:2017rvk, Vorabbi_2018, Vorabbi_2020} is that the polarization
of the target nucleus has to be taken into account. In fact, for a fixed value of the target spin, the nonlocal density obtained from the NCSM method displays a dependence on
the initial and final third component of the target spin.
This difference requires some changes in the formalism and in the derivation of the OP, making the calculations, in particular for targets with high values of the spin, more involved.

The paper is organized as follows: In Sec. II we derive our microscopic OP for non-zero spin nuclei. In Sec. III we discuss relevant details about the nucleon-nucleon ($NN$) chiral potentials employed in the calculations. In Sec. IV we present the results obtained for the differential cross section and the analyzing power of elastic proton-nucleus scattering obtained with our OP  and compare them with the available experimental data. Finally, in Sec. V we summarize our results and draw our conclusions.

\section{Optical Potential for Non-Zero Spin Nuclei}

As we showed in Ref. \cite{Vorabbi:2015nra}, the explicit expression of the optical potential in the impulse approximation can be derived from
the following relation for the elastic $(A + 1)$-body transition operator \cite{PhysRevC.30.1861, PhysRevC.51.1418, PhysRevC.56.2080}
\begin{equation}
T_{\mathrm{el}} = P U P + P U P G_0 (E) T_{\mathrm{el}} \, ,
\end{equation}
where $P$ is conventionally taken as the elastic channel projector, $U$ is the optical potential operator and $G_0(E)$ is the free propagator for the projectile plus target nucleus system.
The elastic OP operator is defined as $U_{\mathrm{el}} \equiv P U P$ and in the impulse approximation it becomes
\begin{equation}\label{masterop}
U_{\mathrm{el}}^{\mathbbmtt{p}} = \sum_{i=1}^Z t_{\mathbbmtt{p} i} + \sum_{i=1}^N t_{\mathbbmtt{p} i} \, ,
\end{equation}
where we explicitly introduced the label $\mathbbmtt{p}$ to denote the projectile and we used $t_{\mathbbmtt{p} i}$ to represent the free two-body scattering matrix of the projectile
and the $i$th nucleon in the target nucleus. It was shown in our previous works that Eq.~(\ref{masterop}) is valid for either protons ($p$) or neutrons ($n$), or even antiprotons ($\bar{p}$).
Even if in this work we will only show results for proton scattering, it is our purpose  to extend the formalism to targets with spin different form zero and we will keep this label in our formalism with the meaning of $\mathbbmtt{p} = (p , n , \bar{p})$.
We denote with ${\bm k}$ and ${\bm k}^{\prime}$ the initial and final momenta of the projectile in the $NA$ frame and we introduce the additional
variables ${\bm q} \equiv {\bm k}^{\prime} - {\bm k}$ (the momentum transfer located along the $\hat{z}$ direction),
${\bm K} \equiv \frac{1}{2} ({\bm k}^{\prime} + {\bm k})$ (the average momentum), ${\bm P}$ as the remainder integration variable~\cite{1997PhDT67W}, and
$\hat{\bm n} \equiv ({\bm K} \times {\bm q}) / |{\bm K} \times {\bm q}|$ (the normal unit vector to the scattering plane).
Working in the momentum representation is a natural choice since the off-shell {\it NN} (or $\bar{p}N$) $t$ matrix is conveniently defined as a function of the relevant momenta.
We also denote with $s$ the spin of the target (here we only treat the elastic scattering so $s$ does not change during the scattering process) and with $\sigma$ and
$\sigma^{\prime}$ the initial and final third component of $s$. To shorten the notation we also define the multi-index $\alpha \equiv ( s , \sigma^{\prime} , \sigma )$ which
contains the target spin quantum numbers.

Using these new variables we can evaluate the operator of Eq.~(\ref{masterop}) in a convenient basis and, after some manipulations, we obtain for the general
matrix element of the OP~\cite{Vorabbi:2015nra}
\begin{widetext}
\begin{equation}\label{nonlocal_op_with_spin}
\begin{split}
U_{\mathrm{el}}^{\mathbbmtt{p}} ({\bm q} , {\bm K} ; \alpha , E) = &\sum_{N =p,n} \int d {\bm P} \; \eta ({\bm q} , {\bm K} , {\bm P}) \;
t_{\mathbbmtt{p} N}^c \left[ {\bm q} , \frac{1}{2} \left( \frac{A+1}{A} {\bm K} + \sqrt{\frac{A-1}{A}} {\bm P} \right) ; E \right] \\
&\times \rho_{\alpha}^{(N)} \left( {\bm P} + \frac{1}{2} \sqrt{\frac{A-1}{A}} {\bm q} , {\bm P} - \frac{1}{2} \sqrt{\frac{A-1}{A}} {\bm q} \right) \\
&+ i ({\bm \sigma} \cdot \hat{\bm n}) \sum_{N =p,n} \int d {\bm P} \; \eta ({\bm q} , {\bm K} , {\bm P}) \;
t_{\mathbbmtt{p} N}^{ls} \left[ {\bm q} , \frac{1}{2} \left( \frac{A+1}{A} {\bm K} + \sqrt{\frac{A-1}{A}} {\bm P} \right) ; E \right] \\
&\times \rho_{\alpha}^{(N)} \left( {\bm P} + \frac{1}{2} \sqrt{\frac{A-1}{A}} {\bm q} , {\bm P} - \frac{1}{2} \sqrt{\frac{A-1}{A}} {\bm q} \right) \, ,
\end{split}
\end{equation}
\end{widetext}
where the first term of the right-hand side is the central part of the OP and the second term is the spin-orbit part, with $i ({\bm \sigma} \cdot \hat{\bm n})$ representing
the spin-orbit operator in momentum space. Here $t_{\mathbbmtt{p} N}^c$ and $t_{\mathbbmtt{p} N}^{ls}$ are the central and the spin-orbit part of the
scattering matrix $t_{\mathbbmtt{p} N}$ and $\eta ({\bm q} , {\bm K} , {\bm P})$ is the Moeller operator included to maintain Lorentz invariance in the transformation from the {\it NA} to
the {\it NN} systems. The OP is also energy dependent and the energy $E$ is fixed at half the kinetic energy of the projectile in the laboratory frame.

An important ingredient of the calculation is the nonlocal density $\rho_{\alpha}^{(N)}$ in momentum space, for which we employ the NCSM
approach~\cite{Navr_til_2009,Barrett:2013nh}. 
The NCSM approach is based on the expansion of the nuclear wave functions in a harmonic oscillator basis and it is thus characterized by the harmonic
oscillator frequency $\hbar \omega$ and the parameter $N_{max}$, which specifies the number of nucleon excitations above the lowest energy configuration allowed by the Pauli principle. 
For all the nuclei considered in this work we used $\hbar \omega =20$ MeV and a $\lambda_{\mathrm{SRG}} = 2.0$ fm$^{-1}$ cutoff for the Similarity Renormalization
Group (SRG) \cite{bogner_2007,roth_2008,jurgenson_2009,bogner_2010} procedure, including the SRG induced three-nucleon ($3N$) force in all the calculations.
For the $N_{max}$ parameter we performed calculations with 8 excitations for $^{9,13}$C and $^{10}$B, 10  for $^{7}$Li and 12 for $^{6}$Li.

The NCSM method is fully self-consistent since center-of-mass contributions have been consistently removed.
In the NCSM approach the one-body nonlocal density is computed in coordinate space and thus it must be transformed to momentum space:
this is done through the Fourier transform and we refer the reader to the Appendix \ref{AppA} for more details.
In momentum space, the general form of the nonlocal density is given in terms of the Jacobi momenta ${\bm \zeta}$ and ${\bm \zeta}^{\prime}$ (see Appendix \ref{AppA}
for the definition)
\begin{equation}\label{general_density}
\begin{split}
\rho_{s^{\prime} \sigma^{\prime} s \, \sigma}^{(N)} ({\bm \zeta}^{\prime} , {\bm \zeta}) = &\frac{1}{\hat{s}^{\prime}} \sum_{K l^{\prime} l}
(s \, \sigma K , \sigma^{\prime} \!-\! \sigma | s^{\prime} \sigma^{\prime}) \, i^{l-l^{\prime}} \\
&\times \rho_{l^{\prime} l}^{(N,K)} (\zeta^{\prime} , \zeta)
{\Big[ Y_{l^{\prime}}^{\ast} (\hat{{\bm \zeta}}^{\prime}) Y_l^{\ast} (\hat{{\bm \zeta}}) \Big]}_{\sigma^{\prime} - \sigma}^{(K)} \, ,
\end{split}
\end{equation}
with the coupled angular functions defined as
\begin{equation}
{\Big[ Y_{l^{\prime}}^{\ast} (\hat{\bm \zeta}^{\prime}) Y_l^{\ast} (\hat{\bm \zeta}) \Big]}_k^{(K)} = \sum_{m^{\prime} m} (l^{\prime} m^{\prime} l m | K k)
Y_{l^{\prime} m^{\prime}}^{\ast} (\hat{\bm \zeta}^{\prime}) \, Y_{l m}^{\ast} (\hat{\bm \zeta}) \, ,
\end{equation}
and with $\rho_{l^{\prime} l}^{(N,K)} (\zeta^{\prime} , \zeta)$ being the radial part of the density. Since in this work we only consider elastic scattering, the spin of the target does not
change during the interaction with the projectile and thus we can set $s = s^{\prime}$ in Eq.~(\ref{general_density}) and drop the dependence on $s^{\prime}$.
In this way we recover the expression for $\rho_{\alpha}^{(N)}$ used in Eq.~(\ref{nonlocal_op_with_spin}).

To compute the OP of Eq.~(\ref{nonlocal_op_with_spin}) we need to interpolate the density and this is done using the relations that connect the Jacobi variables to the
momentum transfer
\begin{equation}\label{relation_zeta_zetap_to_momentum_transfer}
{\bm q} = \sqrt{\frac{A}{A-1}} \big( {\bm \zeta}^{\prime} - {\bm \zeta} \big) \, ,
\end{equation}
and the integration variable
\begin{equation}\label{relation_zeta_zetap_to_integration_variable}
{\bm P} = \frac{1}{2} \big( {\bm \zeta}^{\prime} +  {\bm \zeta} \big) \, .
\end{equation}
We refer the reader to the Appendix \ref{AppB} for more details.
After the calculation of Eq.~(\ref{nonlocal_op_with_spin}), the OP is then interpolated and stored in terms of the variables ${\bm k}$ and ${\bm k}^{\prime}$
(for example see Section II\! C of Ref. \cite{Vorabbi:2015nra}).
Under very general assumptions (i.e. conservation of total angular momentum and parity), the OP is expanded in partial waves as
\begin{equation}\label{op_pw_expansion}
U_{\mathrm{el}}^{\mathbbmtt{p}} ({\bm k}^{\prime},{\bm k}; \alpha , E) = \frac{2}{\pi} \sum_{ljm} \mathcal{Y}_{jm}^{l\frac{1}{2}} ({\hat {\bm k}}^{\prime})
U_{lj}^{\mathbbmtt{p}} (k^{\prime},k; \alpha , E) \mathcal{Y}_{jm}^{l\frac{1}{2}\, \dagger} ({\hat {\bm k}}) \, ,
\end{equation}
where $\mathcal{Y}_{j m}^{l {\scriptstyle \frac{1}{2}}}$ are the usual spin-angular functions defined as
\begin{equation}
\mathcal{Y}_{j m}^{l {\scriptscriptstyle \frac{1}{2}}} ({\hat {\bm k}}) = \sum_{m_l \, m_s} (l \, m_l {\scriptscriptstyle \frac{1}{2}} \, m_s | j m) \,
Y_l^{m_l} ({\hat {\bm k}}) \, \chi_{{\scriptscriptstyle \frac{1}{2}} m_s } \, .
\end{equation}
We immediately see from Eq.~(\ref{op_pw_expansion}) that, for a given value of $s$, the partial wave components of our OP depend on the initial and final
third component of $s$. This is a direct consequence of Eq.~(\ref{general_density}), that enters Eq.~(\ref{nonlocal_op_with_spin}), and thus introduces the dependence
on $\sigma$ and $\sigma^{\prime}$ in the OP. We also notice that the OP of Eq.~(\ref{nonlocal_op_with_spin}) is an operator in the spin space of the projectile only, explaining
the partial wave expansion of Eq.~(\ref{op_pw_expansion}).

Using the same decomposition for the elastic transition operator $T$
\begin{equation}
T_{\mathrm{el}}^{\mathbbmtt{p}} ({\bm k}^{\prime},{\bm k}; \alpha , E) = \frac{2}{\pi} \sum_{ljm} \mathcal{Y}_{jm}^{l\frac{1}{2}} ({\hat {\bm k}}^{\prime})
T_{lj}^{\mathbbmtt{p}} (k^{\prime},k; \alpha , E) \mathcal{Y}_{jm}^{l\frac{1}{2}\, \dagger} ({\hat {\bm k}}) \, ,
\end{equation}
the partial wave components of the resulting transition operator are given by
\begin{widetext}
\begin{equation}\label{transition_amplitude_pw}
T_{lj}^{\mathbbmtt{p}} (k^{\prime},k; \alpha , E) = U_{lj}^{\mathbbmtt{p}} (k^{\prime},k; \alpha , E) + \frac{2}{\pi} \int_0^{\infty} d p \, p^2
\frac{U_{lj}^{\mathbbmtt{p}} (k^{\prime},p; \alpha , E) \, T_{lj}^{\mathbbmtt{p}} (p,k; \alpha , E)}{E - E (p) + i \epsilon} \, .
\end{equation}
\end{widetext}
The scattering amplitude for the elastic scattering of spin $1/2$ projectiles from a target with arbitrary spin $s$ is given by
\begin{equation}\label{general_scattering_amplitude_non_zero_spin_targets}
\begin{split}
f_{\nu^{\prime} \sigma^{\prime} \nu \, \sigma} (\theta) = \: &\delta_{\nu^{\prime} \nu} \, \delta_{\sigma^{\prime} \sigma} \, f_C (\theta) \\
&+ \frac{2}{\pi} \sum_{l j J} (l 0 {\scriptstyle \frac{1}{2}} \nu | j \nu) \, (j \nu s \sigma | J , \nu + \sigma) \\
&\times (l , \nu + \sigma - \nu^{\prime} - \sigma^{\prime}, {\scriptstyle \frac{1}{2}} \nu^{\prime} | j , \nu + \sigma - \sigma^{\prime}) \\
&\times (j , \nu + \sigma - \sigma^{\prime} , s \sigma^{\prime} | J , \nu + \sigma) \\
&\times \sqrt{\frac{2 l + 1}{4 \pi}} \, e^{i 2 \sigma_l (\eta)} \, M_{l j}^{\mathbbmtt{p}} (\alpha , E) \\
&\times Y_l^{\nu+\sigma - \nu^{\prime} - \sigma^{\prime}} (\theta , 0) \, ,
\end{split}
\end{equation}
where the partial wave components of the scattering amplitudes are obtained from the on-shell values of the $T$ matrix as
\begin{equation}\label{scattering_amplitude_pw}
M_{l j}^{\mathbbmtt{p}} (\alpha , E) = - 4 \pi^2 \mu \, T_{lj}^{\mathbbmtt{p}} (k_0 ,k_0 ; \alpha , E) \, ,
\end{equation}
with $k_0$ the on-shell momentum in the $NA$ frame and $\mu$ the reduced mass.
In Eq.~(\ref{general_scattering_amplitude_non_zero_spin_targets}) $\nu$ and $\nu^{\prime}$ represent the initial and final third component of the spin for the spin-1/2 projectile,
$f_C (\theta)$ is the Coulomb scattering amplitude, $\eta$ is the Sommerfeld parameter, and $\sigma_l$ are the Coulomb phase shifts. 

Despite its familiar form,
we notice that Eq.~(\ref{general_scattering_amplitude_non_zero_spin_targets}) differs from the expression that can be found in standard textbooks for two aspects: first,
the partial wave components $M_{l j}^{\mathbbmtt{p}} (\alpha , E)$ do not depend on the total angular momentum $J$ (where $|j - s| \le J \le j+s$) and, second, they depend on the
initial and final third component of the target spin. The first difference derives from how the optical potential is expanded in partial waves (for example see Section II\! C of
Ref. \cite{Vorabbi:2015nra}), while the second one, as explained above, is the direct consequence of the dependence of the target density on $\sigma$ and $\sigma^{\prime}$,
and makes the calculations for targets with high values of spin more involved.
For example, for a spin-3 target we have to calculate Eq.~(\ref{nonlocal_op_with_spin}) 49 times, one for each combination of $\sigma$ and $\sigma^{\prime}$, obtaining
49 different OPs that are then expanded in partial waves and used to solve Eq.~(\ref{transition_amplitude_pw}) to obtain all the
$T_{lj}^{\mathbbmtt{p}} (k_0 ,k_0 ; \alpha , E)$ matrix elements entering Eq.~(\ref{scattering_amplitude_pw}).

To reduce the computational effort, we also mention that we investigated the dependence of our results on the target polarizations.
In particular, since our OP does not contain any spin-orbit term for the target, we can argue that the initial polarization $\sigma$ of the target spin does not change during
the scattering process and, thus, it will be equal to the final one $\sigma^{\prime}$. We explicitly tested this idea performing the calculations for all the nuclei using only
the density components with $\sigma = \sigma^{\prime}$ and setting all the other ones to zero. The results obtained in this way were all matching the full calculations
presented in Sec. IV. This result is very helpful to reduce the computational cost for high values of the target spin: for example, for a spin-3 target we mentioned that we need 49
different OPs to perform the full calculation, while, if we only consider the components of the density with $\sigma = \sigma^{\prime}$, we only need 7 OPs.

From the scattering amplitude we can calculate the differential cross section for an unpolarized beam summing over the final polarizations and averaging over the initial ones
(for example see Ref.~\cite{taylor72})
\begin{equation}\label{general_differential_cross_section_formula}
\frac{d \sigma}{d \Omega} ( \theta ) = \frac{1}{2 (2 s + 1)} \sum_{\nu^{\prime} \sigma^{\prime} \nu \, \sigma}
{\Big| f_{\nu^{\prime} \sigma^{\prime} \nu \, \sigma} (\theta) \Big|}^2 \, .
\end{equation}
In a similar way the analyzing power is obtained as
\begin{equation}\label{general_analyzing_power_formula}
A_y (\theta) = - \frac{\sum_{\sigma^{\prime} \sigma} 2 \, \mathrm{Im} \left[ f_{+{\scriptstyle \frac{1}{2}} \sigma^{\prime} +{\scriptstyle \frac{1}{2}} \sigma } (\theta) \;
f_{-{\scriptstyle \frac{1}{2}} \sigma^{\prime} +{\scriptstyle \frac{1}{2}} \sigma }^{\ast} (\theta) \right]}{\frac{1}{2} \sum_{\nu^{\prime} \sigma^{\prime} \nu \, \sigma}
{\Big| f_{\nu^{\prime} \sigma^{\prime} \nu \, \sigma} (\theta) \Big|}^2} \, .
\end{equation}

The last thing to address is how to include the Coulomb interaction when the projectile is a charged particle.
This is done following the path outlined in Refs.~\cite{PhysRevC.44.1569,elster1993}, which consists in defining a short-range potential
\begin{equation}
\bar{U}_{\mathrm{el}}^{\mathbbmtt{p}} ({\bm k}^{\prime},{\bm k}; \alpha , E) = V_s^C (q)
+ U_{\mathrm{el}}^{\mathbbmtt{p}} ({\bm k}^{\prime},{\bm k}; \alpha , E) \, ,
\end{equation}
obtained from the sum of the Fourier transform of the short-range part of the Coulomb potential in coordinate space $V_s^C (q)$ and the nuclear OP.
The resulting $\bar{U}_{\mathrm{el}}^{\mathbbmtt{p}} ({\bm k}^{\prime},{\bm k}; \alpha , E)$ is then expanded in partial waves and it is transformed
to coordinate space through
\begin{equation}
\begin{split}
\bar{U}_{lj}^{\mathbbmtt{p}} (r^{\prime},r; \alpha , E) = &\frac{4}{\pi^2} \int_0^{\infty} d k^{\prime} \, k^{\prime \, 2} \int_0^{\infty} d k \, k^2 j_l (k^{\prime} r^{\prime}) \\
&\times \bar{U}_{lj}^{\mathbbmtt{p}} (k^{\prime},k; \alpha , E) \, j_l (k r) \, ,
\end{split}
\end{equation}
using the spherical Bessel functions $j_l$, and then it is transformed back to momentum space through
\begin{equation}
\begin{split}
\hat{U}_{lj}^{\mathbbmtt{p}} (k^{\prime},k; \alpha , E) = &\frac{1}{k^{\prime} k} \int_0^{\infty} d r^{\prime} \, r^{\prime} \int_0^{\infty} d r \, r \, F_l (\eta , k^{\prime} r^{\prime}) \\
&\times \bar{U}_{lj}^{\mathbbmtt{p}} (r^{\prime},r; \alpha , E) \, F_l (\eta , k r) \, ,
\end{split}
\end{equation}
using the regular Coulomb functions $F_l$. The partial wave components $\hat{U}_{lj}^{\mathbbmtt{p}} (k^{\prime},k; \alpha , E)$ are then used
to solve Eq.~(\ref{transition_amplitude_pw}) to obtain the $T_{lj}^{\mathbbmtt{p}} (k^{\prime},k; \alpha , E)$ matrix elements.

Finally, we briefly discuss the energy range of applicability of our OP and we try to identify its low- and high-energy limits.

The low-energy limit of the model is dictated by the impulse approximation, introduced to derive
Eq.~(\ref{nonlocal_op_with_spin}). This approximation consists in neglecting the interaction between the struck target nucleon in the target and the residual nucleus.
These effects are very small at 200 MeV and they are negligible at higher energies, however, they become important at energies below 100 MeV, that can be assumed as the
low-energy limit of our model.

The high-energy limit of the model is instead dictated by the applicability of the $NN$ interaction.
The chiral potentials that we are applying (see next section for more details) have a cutoff of 500 MeV/c in terms of the relative momentum.
The equivalent laboratory energy, $T_{\mathrm{lab}} = 2 p^2 / M$, is about 500 MeV. Since the cutoff function in the $NN$ interaction is a Gaussian, it starts acting earlier than 500 MeV/c,
at an energy of $\sim 400$ MeV/c, which is equivalent to $T_{\mathrm{lab}}$ of about 340 MeV. In fact, the phase shifts of $NN$ scattering are perfectly reproduced up to 350 MeV,
that can be taken as the high-energy limit of the model. Concerning the SRG, we notice that our choice, $\lambda_{\mathrm{SRG}} = 2.0$ fm$^{-1}$, is equivalent to about 400 MeV/c,
which then comes down to the same as discussed above.

\section{Chiral Nuclear Potentials}

Before presenting our theoretical predictions we will shortly discuss the relevant details about  the {\it NN} potentials employed for our calculations. In this manuscript we make exclusive use of the most recent generation of {\it NN} chiral potentials derived within the formalism of Chiral Perturbation Theory (ChPT). 
Within this framework, the {\it NN} interaction is governed by the (approximate) chiral symmetry
of the low-energy realization of QCD. As Weinberg suggested a long time ago \cite{WEINBERG1979327}, chiral symmetry greatly constrains construction of the {\it NN} Lagrangian.
In practice, ChPT provides a description of nuclear systems in terms of single and multiple pion exchanges (long- and medium-range components) and contact interactions between the nucleons in order to
parametrize the short-range behavior. For all the details we refer the reader to Refs.~\cite{ Epelbaum:2008ga, Machleidt:2011zz} and to Refs \cite{Machleidt_2021,phillips2021hath,vankolck2021nuclear,entem2021renormalization} for more recent developments and interpretations.
The free parameters of the theory are determined by reproducing data in the $NN$ and $3N$ sector.

In our previous works ~\cite{Vorabbi:2015nra,Vorabbi:2017rvk, Vorabbi_2018, Vorabbi_2020, Gennari:2017yez} we applied chiral {\it NN} potentials at N$^{3}$LO (next-to-next-to-next-to-leading order) \cite{chiralmachleidt_n3lo} and N$^{4}$LO (next-to-next-to-next-to-next-to-leading order) \cite{Entem:2017gor} and for the $3N$ sector at N$^{2}$LO (next-to-next-to-leading order).
At the moment, because of the highly computational resources needed, it is impossible to achieve a full consistency between the {\it NN} potentials employed for the target description and the elastic reaction process, in particular concerning the inclusion of $3N$ forces.
For our calculations we decided to employ the {\it NN} potentials at N$^{3}$LO order \cite{chiralmachleidt_n3lo} for two reasons. On one hand, we have shown in our previous work \cite{Vorabbi:2017rvk, Vorabbi_2018} that including {\it NN} potentials at N$^{4}$LO order does not substantially improve the agreement since the additional contributions are very small. On the other hand, as shown in Ref. \cite{PhysRevC.101.014318}, nuclear structure calculations for light nuclei show that the best agreement with the experimental data is obtained using  the {\it NN} potential at N$^{3}$LO along with $3N$ forces with simultaneous local and nonlocal (3Nlnl) regularization \cite{Navratil2007,Gysbers2019}.

For all the calculations presented in the next Section we used {\it NN} potentials at N$^{3}$LO  \cite{chiralmachleidt_n3lo} with a 500 MeV energy cutoff plus chiral $3N$ forces with low-energy constants $c_D =0.7$,  
$c_E = -0.06$, and $c_i$, taken from Ref. \cite{chiralmachleidt_n3lo}. With the purpose of checking the convergence of our predictions, we also performed a single calculation with {\it NN} potentials at N$^{4}$LO order \cite{Entem:2017gor}, including $3N$ forces
with $c_D = -1.8$, $c_E = -0.31$, and $c_i$ taken from Tab. 9 of Ref. \cite{Entem:2017gor}. 
Since $3N$ forces included in the scattering process must contain {\it medium} corrections, i.e. the presence of a filled Fermi sea \cite{PhysRevC.92.024618, PhysRevC.96.059905, Kohno_ptep_pty001, PhysRevC.98.054617, PhysRevC.102.024611}, we follow here the same procedure outlined in
Refs. \cite{Vorabbi_2020, Holt_2010}, where we varied the density parameter  between 0.08 fm$^{-3}$ and 0.13 fm$^{-3}$.
As a consequence, our results will be drawn as bands and not as single lines, in order to show how much the $3N$ contributions affect the scattering observables varying the matter density.

\section{Results}

In this section we show the theoretical predictions of our OP for the scattering observables of elastic proton scattering off a set of non-zero spin nuclei, with different values of the spin, and compare them to the available experimental data. We have chosen for our investigation a proton energy of about 200 MeV, a value for which the results of our previous work on spin-zero nuclei clearly demonstrated the validity of our microscopic OP.
The experimental data were taken from the experimental nuclear reaction data (EXFOR)~\cite{OTUKA2014272,ZERKIN201831} web utility.

\begin{figure}[t]
\begin{center}
\includegraphics[scale=0.34]{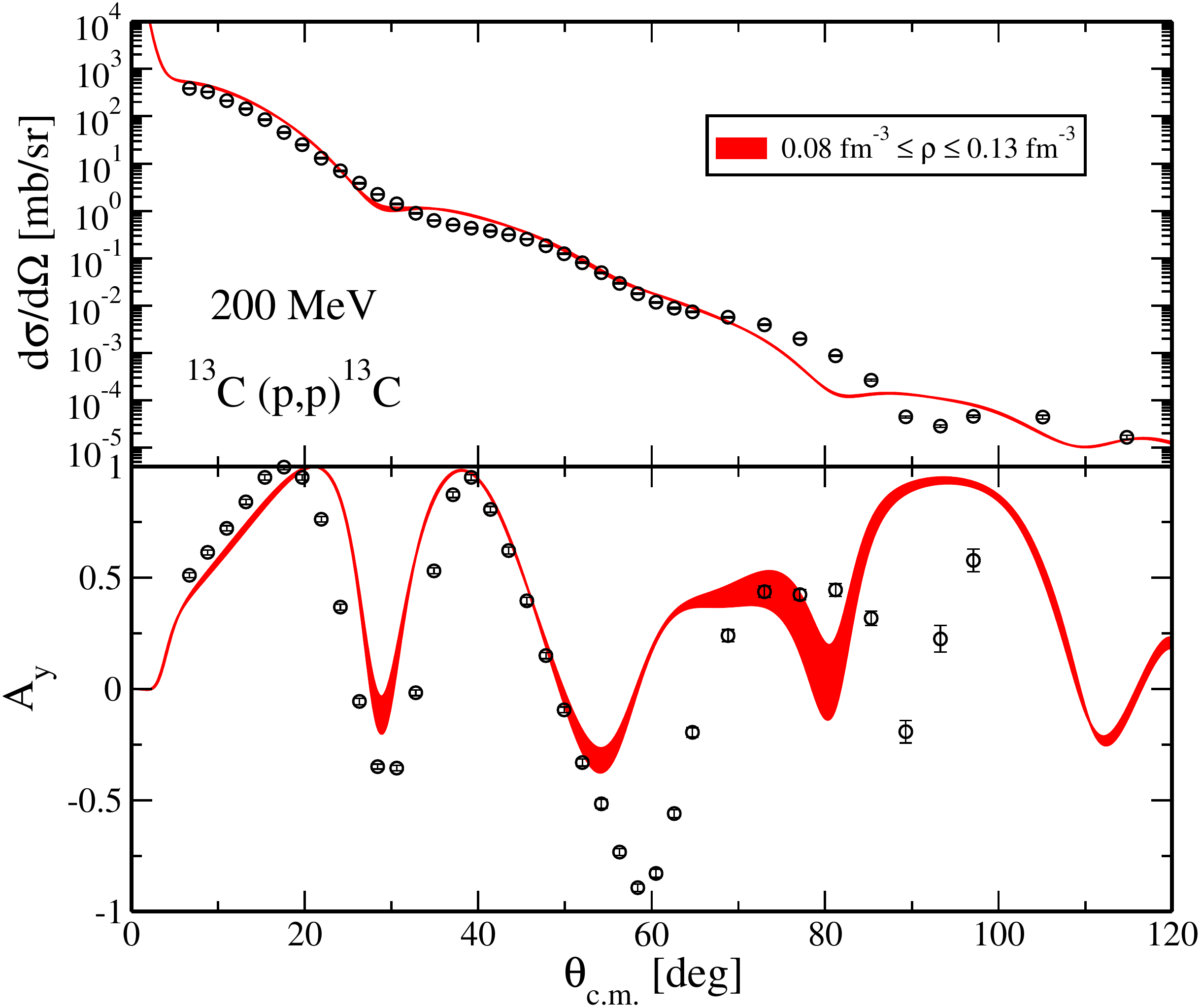}
\caption{ (Color online)  Differential cross section (upper panel) and analyzing power (lower panel), as functions of the center-of-mass scattering angle, for 200 MeV protons elastically scattered from $^{13}$C ($J^{\pi} = {1/2}^-$). The results were obtained using Eq.~(\ref{nonlocal_op_with_spin}), where 
the {\it NN} $t$ matrix is computed with the {\it NN} chiral interaction at N$^3$LO order of Ref. \cite{chiralmachleidt_n3lo}, supplemented by a density dependent {\it NN} interaction (where the baryon density is varied in the range between 0.08 fm$^{-3}$ and 0.13 fm$^{-3}$) and the one-body nonlocal density matrices computed with the NCSM method using {\it NN} \cite{chiralmachleidt_n3lo} and 3Nlnl \cite{Navratil2007,Gysbers2019} chiral interactions.
Experimental data from Ref. \cite{Meyer:1981na}. \label{figC13}}
\end{center}
\end{figure}

As a first case we consider elastic proton scattering on $^{13}$C target. The ground state of $^{13}$C has spin and parity quantum number 
$J^\pi = 1/2^-$ and it is therefore well suited to test our theoretical approach.

Measurements were carried out  using the polarized proton beam from the Indiana University Cyclotron Facility with 200 MeV of mean energy (the mean scattering energy was varied between 199.8 and 200.1 MeV) \cite{Meyer:1981na}. Differential cross sections and analyzing powers were measured for two isotopes of carbon: $^{12}$C and $^{13}$C. 
The comparison of our previous results with $^{12}$C ($J^{\pi} = 0^+$) data can be found  in Figs. 5 and 6 of Ref.~\cite{Vorabbi:2020cgf}. The comparison with the empirical data for $^{13}$C  \cite{Meyer:1981na} is shown in Fig.~\ref{figC13}, where the calculated differential cross section $d\sigma/d\Omega$ and analyzing power $A_y$ are displayed as functions of the center-of-mass scattering angle $\theta_{c.m.}$. 
As previously mentioned, in order to check the effects of $3N$ contributions, we let the density parameter of the effective $3N$ forces vary in a reasonable range for the matter density: 0.08 fm$^{-3}$ $\le \rho \le$ 0.13 fm$^{-3}$. The effects of genuine $3N$ forces turn out to be rather small for the differential cross section, where the thin thickness of the band indicates that  the results obtained with different values of the density parameter are basically on top of each other, and just a little bit larger for the polarization observable $A_y$. 
The fact that the effects are not larger than those obtained in our previous work on spin-zero nuclei \cite{Vorabbi_2020} could be due to the fact that the two-body approximation of the $3N$ forces at N$^{2}$LO order is performed in the approximation of symmetric spin-saturated nuclear matter. 

Generally speaking, the agreement with the empirical data shown in Fig. \ref{figC13} is quite satisfactory, especially if we consider that no adjustments of the OP have been made, since our OP derivation is fully microscopic. We see reasonable agreement with the data for scattering angles up to $\sim 70^{\rm o}$ for the cross section and $\sim 55^{\rm o}$ for the analyzing power, after which, with increasing scattering angle, the agreement worsens. The overall agreement between the results of our microscopic OP and the empirical data is of about the same quality as that obtained in Ref.~\cite{Vorabbi_2020} for $^{12}$C. 

\begin{figure}[t]
\begin{center}
\includegraphics[scale=0.34]{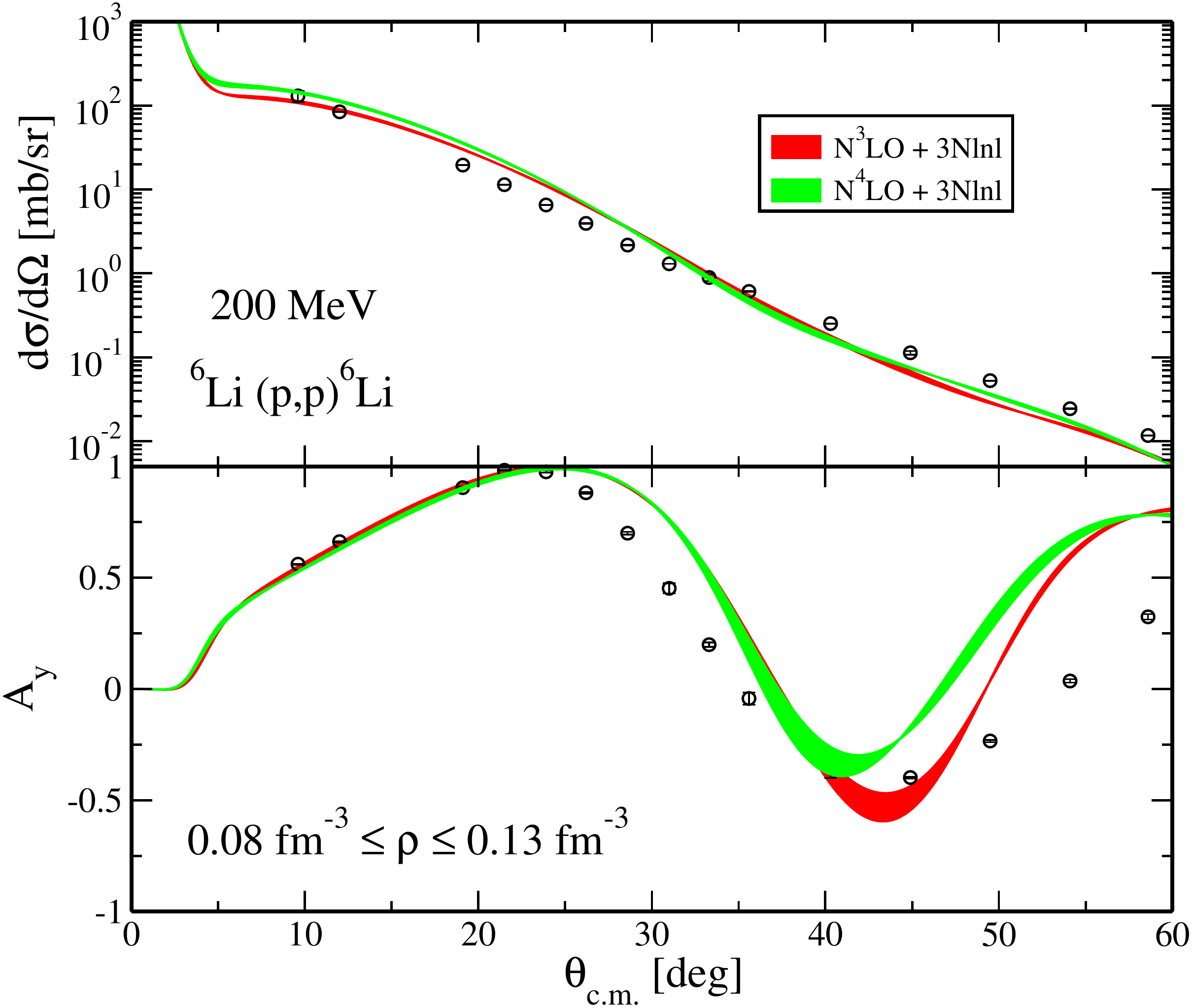}
\caption{ (Color online) The red band is the same as in Fig.~\ref{figC13} but for $^6$Li ($J^{\pi} = {1}^+$) at 200 MeV. The green band is the corresponding result at N$^4$LO order \cite{Entem:2017gor} of the chiral expansion. Experimental data from Ref. \cite{Glover:1990zz}. 
\label{figLi6} }
\end{center}
\end{figure}

\begin{figure}[t]
\begin{center}
\includegraphics[scale=0.34]{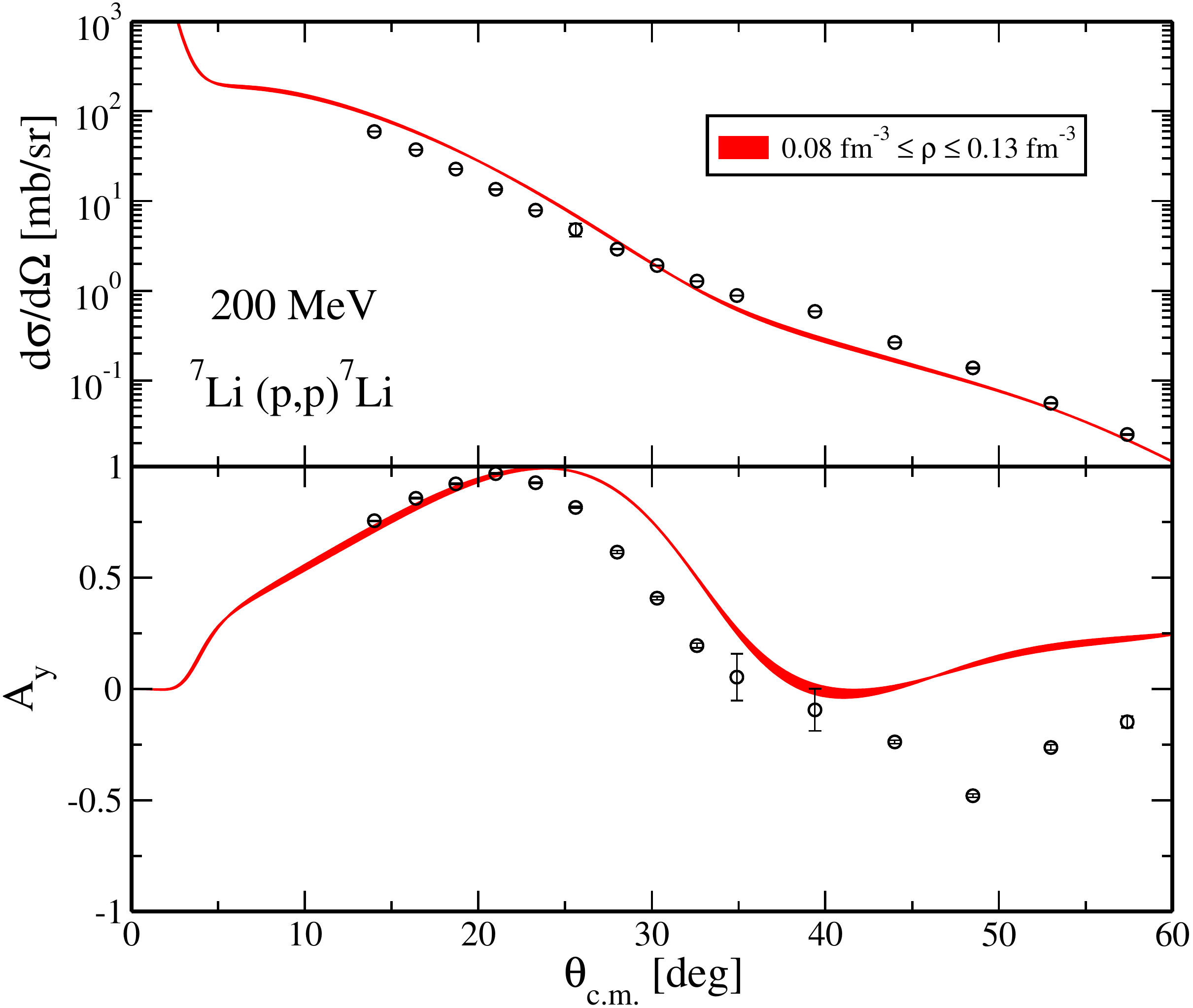}
\caption{ (Color online) The same as in Fig.~\ref{figC13} but for $^7$Li($J^{\pi} = {3/2}^-$) at 200 MeV. Experimental data from Ref. \cite{Glover:1991zz}. 
\label{figLi7} }
\end{center}
\end{figure}

The calculated differential cross section and the analyzing power for elastic proton scattering on $^{6}$Li target ($J^{\pi} = 1^+$) are displayed in 
Fig.~\ref{figLi6} as functions of the center-of-mass scattering angle $\theta_{c.m.}$. 
The experimental data were measured at the Indiana University Cyclotron Facility using a polarized proton beam at a laboratory
bombarding energy of 200.4 MeV \cite{Glover:1990zz}. 
With the purpose of checking the convergence of the theoretical predictions we compare in the figure  the results obtained  with {\it NN} potentials at N$^{3}$LO (red bands) and N$^{4}$LO order (green bands). We can see from the figure that the differences between the two results are small, practically negligible for the cross section and somewhat larger for $A_y$, where, as expected, the results at  N$^{3}$LO give a better agreement with the experimental data. 
Also in the case of $^{6}$Li the agreement of the results with the empirical data is satisfactory and the effects of genuine $3N$ forces  turn out to be rather small for the differential cross section and a little bit larger for the analyzing power.

The results for  $^{7}$Li ($3/2^{-}$) are presented in Fig.~\ref{figLi7}.
The differential cross section and analyzing power were measured at the Indiana University Cyclotron Facility using a polarized proton beam at a laboratory
bombarding energy of 200.4 MeV \cite{Glover:1991zz}.
Also in this case the agreement between the theoretical prediction and the empirical data is good for the differential cross section, over all the angular distribution shown in the figure, and satisfactory for the analyzing power for values of the scattering angle up to $\sim 45^{\rm o}$.
The effects of genuine $3N$ forces turn out to be generally rather small.  

\begin{figure}[t]
\begin{center}
\includegraphics[scale=0.34]{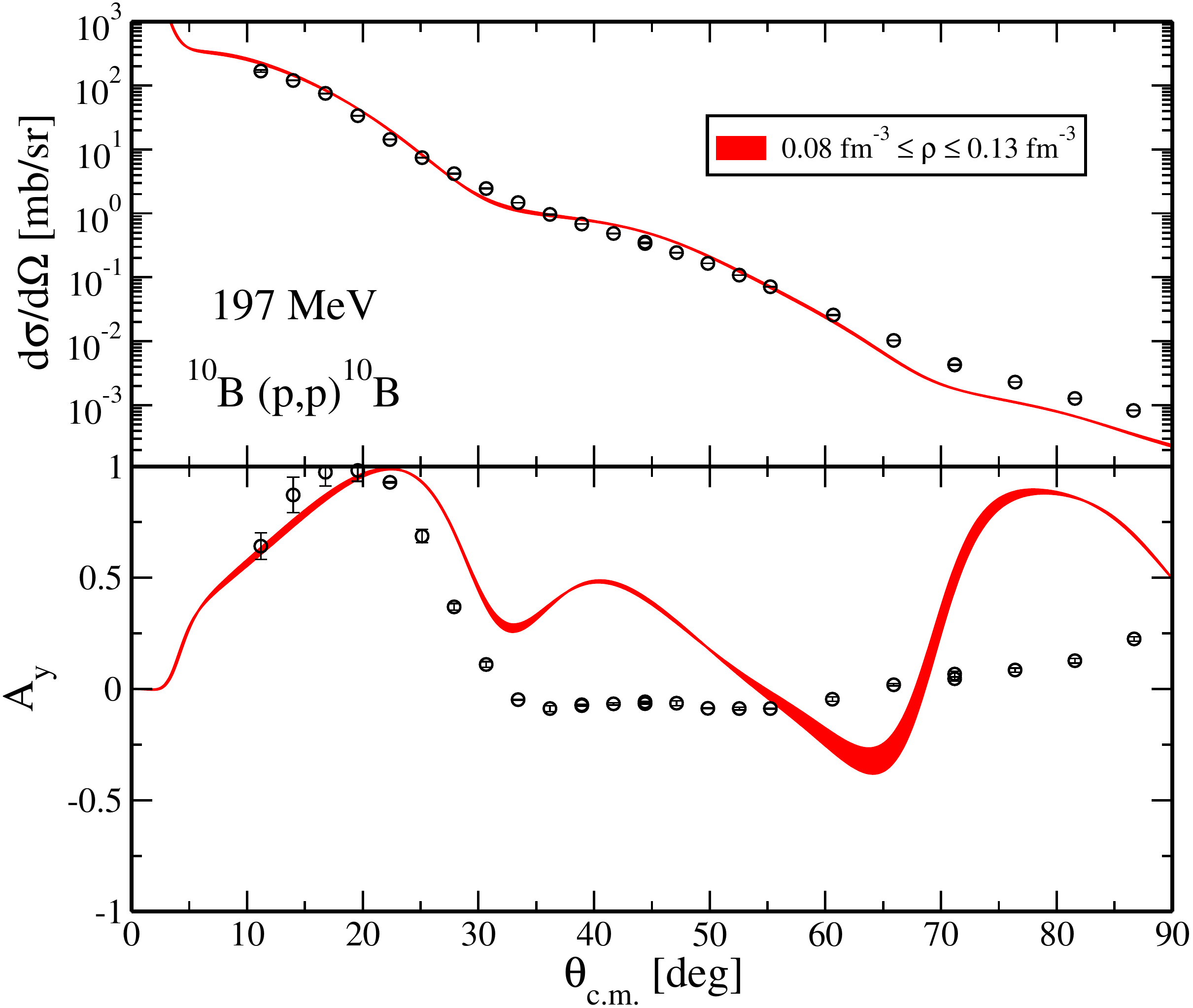}
\caption{ (Color online)  The same as in Fig. \ref{figC13} but for $^{10}$B ($J^{\pi} = {3}^+$) at 197 MeV. Empirical data from Ref. \cite{Betker:2005kt}.
\label{figB10}}
\end{center}
\end{figure}

In Fig.~\ref{figB10} we show our results for $^{10}$B, a nucleus with a high value of the ground-state spin $J^{\pi} = {3}^+$.
With increasing spin value, our calculations become more involved.
The differential cross section and analyzing power were measured for 197 MeV proton scattering at the Indiana University Cyclotron Facility \cite{Betker:2005kt}.
Considering the high value of the spin, the agreement between our theoretical predictions and the empirical data is satisfactory. This is in contradiction to what is often stated in the literature, for instance, in Ref. \cite{Betker:2005kt}, that addressing the elastic scattering data with only optical model techniques would lead to significant problems with the quality of the agreement to any particular portion of the data. In Fig.~\ref{figB10} the experimental cross section is overall well described, although somewhat underpredicted for the highest values of the scattering angle. The agreement is worse for the analyzing power, where our results are able to describe the data only for the lowest values of the scattering angle, up to $\sim 20^{\rm o}-30^{\rm o}$.

The agreement of our results with the experimental data is always worse for the analyzing power than for the cross section.
This is in general true also for the spin-zero targets treated in our previous works, and it is not surprising, since the analyzing power is more sensitive and thus more difficult to reproduce.
In general, from our previous works, all performed for targets with spin equal to zero, we saw that the model is able to provide a result for the $A_y$ that describes the general shape of the
data but the minima are never deep enough to provide a good description of the data. The cases treated in this work are even more complicated, because the final result for the $A_y$
is an average between the analyzing powers obtained for all the specific combinations of $\sigma$ and $\sigma^{\prime}$, as shown in Eq.~(\ref{general_analyzing_power_formula}).
At the current stage, it is not clear how to improve the results for this observable. One possibility is to include in the model the second-order term of the spectator expansion, which is
feasible in principle, but it is complicated and represents a future challenge. Another possibility is to change the $NN$ interaction, following for example
Refs.~\cite{Burrows:2018ggt, Burrows:2020qvu}, where the N$^2$LO$_{\mathrm{opt}}$ potential~\cite{n2lo_opt} was used to construct an OP for spin-zero targets
which provided a very good agreement between the theoretical calculations and the experimental data of the $A_y$.

\begin{figure}[t]
\begin{center}
\includegraphics[scale=0.34]{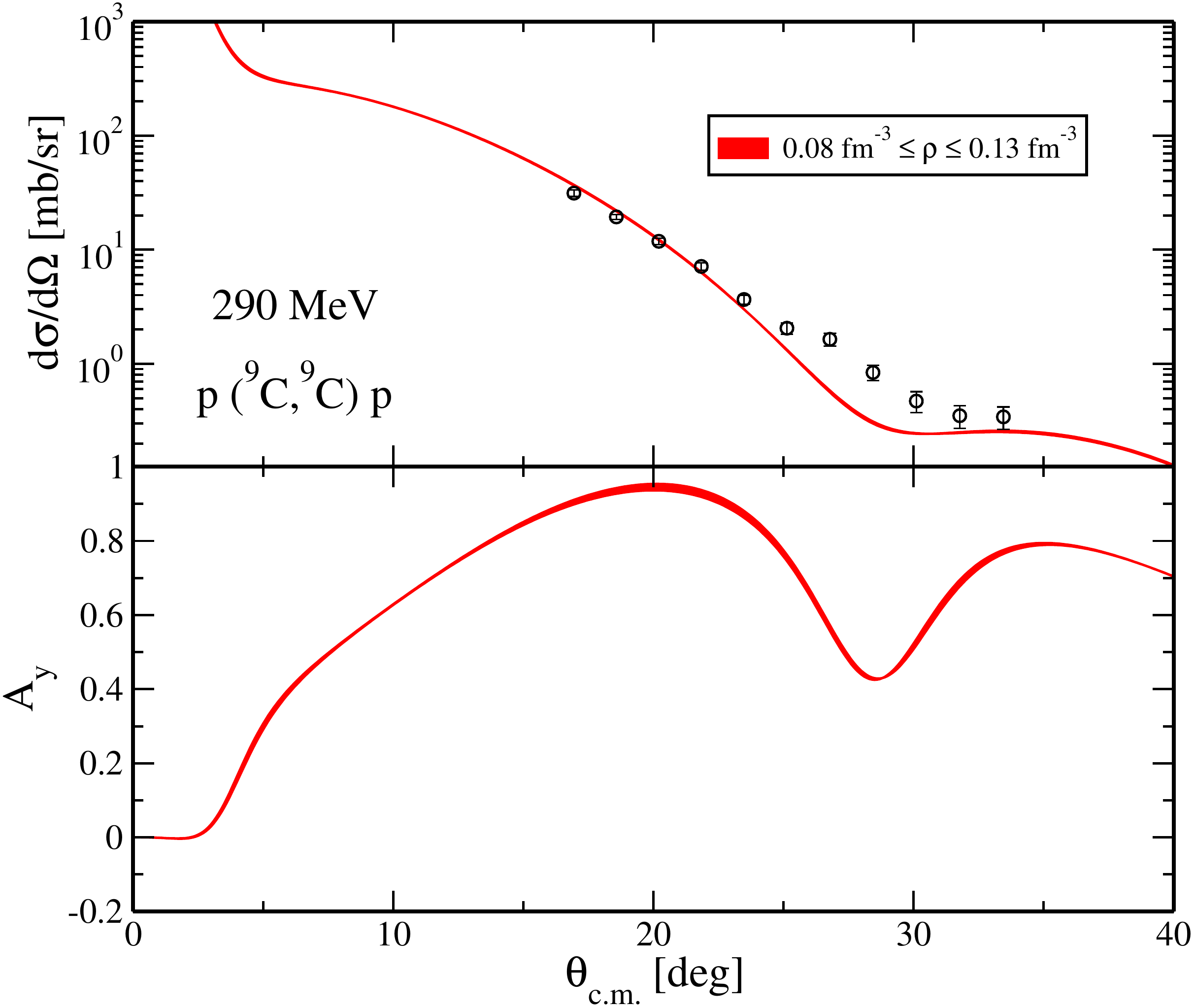}
\caption{ (Color online) Differential cross section (upper panel) and analyzing power (lower panel) for the H($^9$C,p) reaction  at 290 MeV/nucleon, as functions of the center-of-mass scattering angle. The ground state of $^{9}$C has spin and parity quantum number $J^\pi = 3/2^-$.
The results were obtained using the same conditions reported in Fig. \ref{figC13}. Empirical data from Ref. \cite{PhysRevC.87.034614}. \label{figC9}}
\end{center}
\end{figure}

A last example is presented in Fig.~\ref{figC9}, where the results of calculations performed for elastic proton scattering from $^9$C ($J^{\pi} = {3/2}^-$) with a 290 MeV/nucleon $^9$C, in inverse kinematics configuration, are displayed and compared with the available empirical data. 
The experiment was performed at a secondary beam course in the Heavy Ion Medical Accelerator in Chiba (HIMAC) of the National Institute of Radiological Science (NIRS), where the angular distribution of the differential cross section of the H($^9$C,p) reaction at 277-300 MeV/nucleon was measured with a newly designed recoil proton spectrometer \cite{PhysRevC.87.034614}.
In this case empirical data are available only for the cross section, which is reasonably well described by our theoretical predictions. The effects of genuine $3N$ forces are small for the differential cross section and just a little bit larger for $A_y$.

\section{Summary and Conclusions}

In a series of papers, over the last few years, we constructed a microscopic optical potential for elastic (anti)nucleon-nucleus scattering  from chiral potentials.
The OP was derived at first order of the spectator expansion in Watson multiple scattering theory and its final expression is a folding integral between the $NN$ $t$ matrix and the nuclear density of the target. In the calculations, $NN$ and $3N$ chiral interactions are used for the target density, while for the $t$ matrix the effect of the $3N$ interaction is approximated with a density dependent $NN$ interaction obtained from averaging over the Fermi sphere. Our OP was successfully tested in comparison with experimental data, where it is able to provide a reasonably good description of the experimental cross section and polarization observables of different nuclei. However, till now, it was applied only to spin-zero nuclei. 

In the present work we have extended our microscopic OP to non-zero spin target nuclei.
The extension requires some changes in the derivation of the OP and in the formalism. The main difference with respect  to the zero-spin case is that now the target density displays
an additional dependence on the initial and final third component of the target spin, which is then propagated to the OP.
This difference makes the calculations more and more involved and time consuming with the increasing value of the target spin.

Theoretical predictions for the cross section and the analyzing power of elastic proton scattering off a set of nuclei with different values of the spin in their ground state (between $J= 1/2$ and $3$) have been presented and discussed in comparison with the available data, for a proton energy of about 200 MeV.

We checked  the convergence of the theoretical predictions with a single example for elastic proton scattering off $^6$Li, comparing the results obtained  with {\it NN} potentials at N$^{3}$LO and N$^{4}$LO order. The differences between the two results are small, practically negligible for the cross section and somewhat larger for the analyzing power, where the results at N$^{3}$LO give a better agreement with the experimental data. A better agreement of the result with the potential at N$^{3}$LO could be expected, since from nuclear structure calculations for light nuclei \cite{PhysRevC.101.014318} it was shown that the best agreement with the experimental data is obtained using the {\it NN} potential at N$^{3}$LO. Therefore we decided to use the {\it NN} potential at N$^{3}$LO as a basis for our calculations.

In order to test the validity of our microscopic OP when extended to non-zero spin nuclei, we have compared results obtained for a set of targets with different values of the spin. As the target's spin value increases, the calculations become more and more involved, but our results generally give remarkably equivalent agreement for all considered values of the spin.
The quality of the agreement is comparable to the one obtained in our previous work for spin-zero nuclei at the same energy around 200 MeV.
The experimental differential cross sections are in general well described by our theoretical predictions, while the description of the analyzing power is less satisfactory. The agreement between the results of the calculations and the empirical data gets worse as the scattering angle increases, as it was also found in our previous work for spin-zero nuclei.

The effects of genuine $3N$ forces turn out to be rather small for the differential cross section and just a little bit larger for the analyzing power. The fact that these effects are not larger than those obtained for spin-zero nuclei could be due to the fact that the two-body approximation of the $3N$ forces at N$^2$LO is performed in the approximation of symmetric spin-saturated nuclear matter.

We also performed calculations of elastic proton scattering off $^{9}$C, which was measured in an inverse kinematics configuration.
Our theoretical predictions are able to give an overall good  description of the experimental differential cross section of the H($^9$C,p) reaction at 290 MeV/nucleon. 
The inverse kinematics configuration is necessary to study exotic nuclei that have very short average life times. We note that just for the study of exotic nuclei a microscopic approach to the OP should be preferable to a phenomenological one, since in situations for which empirical data are not yet available or are still scarce an OP better founded on theoretical grounds should be able to give more reliable predictions and to assess the impact of the adopted approximations.  

Our results show that our microscopic OP is able to give a remarkable agreement with the experimental data also on nuclear targets with spin different from zero: the extension of the OP to nuclei with $J^\pi \ne 0^+$ is well under control. This is an important achievement in itself, which allows us to apply our OP to a wider range of cases. This is, however, also an important step forward towards the extension of the OP to inelastic {\it NA} scattering, which will be our next goal.
Recent studies have tried to use experiments in inverse  kinematics with the purpose to determine the density of matter of nuclear systems.
However, these measurements are not free from sizable uncertainties and it becomes important to establish how effectively the elastic scattering of protons is related to the density of nuclear matter. In the data analysis of these experiments an essential step of the procedure is the subtraction of the inelastic contributions. In this perspective, if we want to establish a consistent microscopic approach for inelastic $NA$ scattering, it is mandatory to test the microscopic OP potential even on states with spin-parity quantum numbers $J^\pi \ne 0^+$.
With the present work, showing that the derivation of optical potentials between states 
with J$^\pi \ne 0^+$ is completely under control, we paved the way toward a full microscopic approach to inelastic {\it NA}  scattering.

%%%%%%%%%%%%%%%%%%%%%%%%% Acknowledgements %%%%%%%%%%%%%%%%%%%%%%%%%

\section{Acknowledgements}

The work at Brookhaven National Laboratory was sponsored by the Office of Nuclear Physics, Office of Science of the U.S. Department of Energy under
Contract No. DE-AC02-98CH10886 with Brookhaven Science Associates, LLC.
The work at TRIUMF was supported by the NSERC Grant No. SAPIN-2016-00033. TRIUMF receives federal funding via a contribution agreement with the National
Research Council of Canada.
Computing support came from an INCITE Award on the Summit supercomputer of the Oak Ridge Leadership Computing Facility (OLCF) at ORNL, from Westgrid and
Compute Canada.
The work by R.M.\ was supported in part by the U.S. Department of Energy
under Grant No.~DE-FG02-03ER41270.

%%%%%%%%%%%%%%%%%%%%%%%%%%%% Appendix %%%%%%%%%%%%%%%%%%%%%%%%%%%

\begin{appendix}

\section{Nonlocal density in momentum space}
\label{AppA}

In our approach, the one-body nonlocal density is computed in coordinate space using the Jacobi coordinates ${\bm \xi}$ and ${\bm \xi}^{\prime}$, and its general form is
given by ($N = n, p$)
\begin{equation}
\begin{split}
\rho_{s^{\prime} \sigma^{\prime} s \, \sigma}^{(N)} ({\bm \xi}^{\prime} , {\bm \xi}) = &\frac{1}{\hat{s}^{\prime}} \sum_{K l^{\prime} l}
(s \, \sigma K , \sigma^{\prime} \!-\! \sigma | s^{\prime} \sigma^{\prime}) \, \rho_{l^{\prime} l}^{(N,K)} (\xi^{\prime} , \xi) \\
&\times {\Big[ Y_{l^{\prime}}^{\ast} (\hat{{\bm \xi}}^{\prime}) Y_l^{\ast} (\hat{{\bm \xi}}) \Big]}_{\sigma^{\prime} - \sigma}^{(K)}  \, ,
\end{split}
\end{equation}
where
\begin{equation}
{\Big[ Y_{l^{\prime}}^{\ast} (\hat{\bm \xi}^{\prime}) Y_l^{\ast} (\hat{\bm \xi}) \Big]}_k^{(K)} = \sum_{m^{\prime} m} (l^{\prime} m^{\prime} l m | K k) \,
Y_{l^{\prime} m^{\prime}}^{\ast} (\hat{\bm \xi}^{\prime}) \, Y_{l m}^{\ast} (\hat{\bm \xi}) \, ,
\end{equation}
and $\rho_{l^{\prime} l}^{(N,K)} (\xi^{\prime} , \xi)$ is the radial part of the non local density.
In our convention, the Jacobi variables ${\bm \xi}$ and ${\bm \xi}^{\prime}$ are both defined as
\begin{equation}
{\bm \xi} = \sqrt{\frac{A-1}{A}} \left[ \frac{1}{A-1} \sum_{i=1}^{A-1} {\bm r}_i - {\bm r}_{A} \right] \, ,
\end{equation}
where ${\bm r}_i$ is the coordinate of the $i$th nucleon in the target nucleus.
The double Fourier transform to momentum space is given by the following relation
\begin{equation}\label{double_fourier_transform_to_momentum_space}
\rho ({\bm \zeta}^{\prime} , {\bm \zeta}) = \int d^3 \xi^{\prime} \int d^3 \xi \braket{{\bm \zeta}^{\prime}|{\bm \xi}^{\prime}}
\rho ({\bm \xi}^{\prime},{\bm \xi}) \braket{{\bm \xi} | {\bm \zeta}} \, ,
\end{equation}
where
\begin{align}
\braket{{\bm \zeta}^{\prime}|{\bm \xi}^{\prime}}
&= \sqrt{\frac{2}{\pi}} \sum_{l m} {(-i)}^l j_l (\zeta^{\prime} \xi^{\prime}) Y_{l m}^{\ast} (\hat{\bm \zeta}^{\prime}) Y_{l m} (\hat{\bm \xi}^{\prime}) \, , \label{plane_wave_expansion_1} \\
\braket{{\bm \xi}|{\bm \zeta}}
&= \sqrt{\frac{2}{\pi}} \sum_{l m} i^l j_l (\zeta \xi ) Y_{l m} (\hat{\bm \xi}) Y_{l m}^{\ast} (\hat{\bm \zeta}) \label{plane_wave_expansion_2} \, .
\end{align}
In the previous expressions we introduced the Jacobi momenta ${\bm \zeta}$ and ${\bm \zeta}^{\prime}$, that in our convention are both defined as
\begin{equation}
{\bm \zeta} = \sqrt{\frac{A-1}{A}} \left[ \frac{1}{A-1} \sum_{i=1}^{A-1} {\bm k}_i - {\bm k}_{A} \right] \, ,
\end{equation}
where ${\bm k}_i$ is the momentum of the $i$th nucleon in the target nucleus.
Inserting Eqs. (\ref{plane_wave_expansion_1}) and (\ref{plane_wave_expansion_2}) into Eq. (\ref{double_fourier_transform_to_momentum_space}),
the nonlocal density in momentum space is expressed as
\begin{equation}\label{general_form_3d_nonlocal_dens}
\begin{split}
\rho_{s^{\prime} \sigma^{\prime} s \, \sigma}^{(N)} ({\bm \zeta}^{\prime} , {\bm \zeta}) = &\frac{1}{\hat{s}^{\prime}} \sum_{K l^{\prime} l}
(s \, \sigma K , \sigma^{\prime} \!-\! \sigma | s^{\prime} \sigma^{\prime}) \, i^{l-l^{\prime}} \\
&\times \rho_{l^{\prime} l}^{(N,K)} (\zeta^{\prime} , \zeta) {\Big[ Y_{l^{\prime}}^{\ast} (\hat{{\bm \zeta}}^{\prime}) Y_l^{\ast} (\hat{{\bm \zeta}}) \Big]}_{\sigma^{\prime} - \sigma}^{(K)} \, ,
\end{split}
\end{equation}
with the angular part given by
\begin{equation}\label{general_angular_part_nonloc_dens}
{\Big[ Y_{l^{\prime}}^{\ast} (\hat{\bm \zeta}^{\prime}) Y_l^{\ast} (\hat{\bm \zeta}) \Big]}_k^{(K)} = \sum_{m^{\prime} m} (l^{\prime} m^{\prime} l m | K k) \,
Y_{l^{\prime} m^{\prime}}^{\ast} (\hat{\bm \zeta}^{\prime}) \, Y_{l m}^{\ast} (\hat{\bm \zeta}) \, ,
\end{equation}
and the radial component defined as
\begin{equation}
\begin{split}
\rho_{l^{\prime} l}^{(N,K)} (\zeta^{\prime} , \zeta) \equiv &\frac{2}{\pi} \int_0^{\infty} d \xi^{\prime} \xi^{\prime \, 2} \int_0^{\infty} d \xi \xi^2 \, j_{l^{\prime}} (\zeta^{\prime} \xi^{\prime}) \\
&\times \rho_{l^{\prime} l}^{(N,K)} (\xi^{\prime} , \xi) \, j_l (\zeta \xi) \, .
\end{split}
\end{equation}

\section{Interpolation of the nonlocal density in momentum space}
\label{AppB}

We see from Eq.~(\ref{general_form_3d_nonlocal_dens}) that the nonlocal density is expressed in momentum space using the variables ${\bm \zeta}^{\prime}$
and ${\bm \zeta}$, but the calculation of Eq.~(\ref{nonlocal_op_with_spin}) requires the knowledge of the density in terms of the variables ${\bm q}$ and ${\bm P}$.
In general, the density is first computed in momentum space using the variables $\zeta^{\prime}$, $\zeta$, and $\cos \gamma$, and then it is interpolated and stored in
terms of $q$, $P$, and $\cos \theta_P$. Here we use $\gamma$ to represent the angle between ${\bm \zeta}^{\prime}$ and ${\bm \zeta}$, and $\theta_P$ to represent the angle
between ${\bm q}$ and ${\bm P}$. This procedure was used in all our previous works on zero-spin nuclei, where $K$ can only assume the value zero.
However, when the nucleus spin is different from zero, we have that $0 \le K \le 2 s$ ($s$ is the target spin) and this procedure does not work anymore.
In fact, from Eq.~(\ref{general_angular_part_nonloc_dens}) we see that, except for $K=0$, it is not possible to use the addition theorem of the spherical harmonics and thus
reduce the angular part evaluation to a function of $l$ and $\cos \gamma$. Thus, we need to develop a different method to perform the interpolation.

Our goal is to express the density as a function of the variables ${\bm q}$ and ${\bm P}$, more precisely, as $\rho (q , P , \cos \theta_P)$.
Keeping in mind that ${\bm q}$ is located along the $\hat{z}$ axis, we can start writing the Cartesian components of ${\bm q}$ and ${\bm P}$ as
\begin{align}
{\bm q} &= (0 , 0 , q)^{\mathrm{T}} \, , \\
{\bm P} &= (P \sin \theta_P \cos \phi_P , P \sin \theta_P \sin \phi_P , P \cos \theta_P )^{\mathrm{T}} \, ,
\end{align}
and from Eqs. (\ref{relation_zeta_zetap_to_momentum_transfer}) and (\ref{relation_zeta_zetap_to_integration_variable}), which relate the set of variables
$({\bm q} , {\bm P})$ to $({\bm \zeta}^{\prime} , {\bm \zeta})$, we can calculate the Cartesian components of the last set of variables, obtaining
\begin{equation}
{\bm \zeta}^{\prime} =
\begin{pmatrix}
P \sin \theta_P \cos \phi_P \\
P \sin \theta_P \sin \phi_P \\
P \cos \theta_P + \frac{1}{2} \sqrt{\frac{A-1}{A}} q
\end{pmatrix}
 \, ,
\end{equation}
and
\begin{equation}
{\bm \zeta} =
\begin{pmatrix}
P \sin \theta_P \cos \phi_P \\
P \sin \theta_P \sin \phi_P \\
P \cos \theta_P - \frac{1}{2} \sqrt{\frac{A-1}{A}} q 
\end{pmatrix}
\, .
\end{equation}
The spherical components of ${\bm \zeta}^{\prime} = (\zeta^{\prime} , \theta^{\prime} , \phi^{\prime})$ are then obtained as
\begin{align}
\zeta^{\prime} &= \sqrt{P^2 + \frac{A-1}{4 A}q^2 + \sqrt{\frac{A-1}{A}} q P \cos \theta_P} \, , \label{interpolation_1} \\
\cos \theta^{\prime} &= \frac{P \cos \theta_P + \frac{1}{2} \sqrt{\frac{A-1}{A}} q}{\zeta^{\prime}} \, , \label{interpolation_2} \\
\phi^{\prime} &= \phi_P \, ,
\end{align}
while for ${\bm \zeta} = (\zeta , \theta , \phi )$ we have
\begin{align}
\zeta &= \sqrt{P^2 + \frac{A-1}{4 A}q^2 - \sqrt{\frac{A-1}{A}} q P \cos \theta_P} \, , \label{interpolation_3} \\
\cos \theta &= \frac{P \cos \theta_P - \frac{1}{2} \sqrt{\frac{A-1}{A}} q}{\zeta} \, , \label{interpolation_4} \\
\phi &= \phi_P \, .
\end{align}
From these results we see that the interpolation does not depend on $\phi_P$, thus, the desired density can be obtained evaluating Eq.~(\ref{general_form_3d_nonlocal_dens})
using the set of variables $(\zeta^{\prime}, \zeta, \cos \theta^{\prime}, \cos \theta)$ with $\phi^{\prime} = \phi = 0$. Then, we use Eqs. (\ref{interpolation_1}), (\ref{interpolation_2}),
(\ref{interpolation_3}), and (\ref{interpolation_4}) to interpolate and store the density in terms of $q$, $P$, and $\cos \theta_P$.

\end{appendix}

%\bibliography{biblio}

\end{document}